\documentclass[11pt]{article}

\usepackage{graphicx}
\usepackage{rotating}
\usepackage{hyperref}
\usepackage{cite}
\usepackage[all]{hypcap}

\hypersetup{
    bookmarks=true,         
    unicode=false,          
    pdffitwindow=true,      
    pdftitle={Information Resources in High-Energy Physics},    
    pdfauthor={Anne Gentil-Beccot et al.},     
    pdfnewwindow=true,      
    pdfkeywords={keywords}, 
    colorlinks=true,       
    linkcolor=red,          
    citecolor=red,        
    urlcolor=blue           
}

\begin{document}
\newcommand{\etal}{{\it et~al.}}

\begin{flushright}
{
CERN--OPEN--2008--010\\
DESY--08--040\\
FERMILAB-PUB-08-077-BSS\\
SLAC--PUB--13199\\
April 7, 2008\\
}
\end{flushright}
\begin{center}
{\Large\bf 
Information Resources in High-Energy Physics\\
}
{\bf Surveying the Present Landscape and Charting the Future Course}

\vspace{0.5cm}

Anne Gentil-Beccot, Salvatore Mele \\
CERN, European Organization for Nuclear Research\\
CH1211, Gen\`eve 23, Switzerland\\
\vspace{0.3cm}
Annette Holtkamp\\
Deutsches Elektronen-Synchrotron DESY\\
Notkestra\ss{}e 85, D-22607 Hamburg, Germany\\
\vspace{0.3cm}
Heath B. O'Connell\\
Fermilab Library MS 109\footnote{Operated by Fermi Research Alliance, LLC under Contract No. 
DE-AC02-07CH11359 with the United States Department of Energy}\\
P.O. Box 500, Batavia, Illinois 60510 USA\\
\vspace{0.3cm}
Travis C. Brooks \\
SLAC, Stanford Linear Accelerator Center Library/SPIRES Databases\footnote{Work partly supported by Department of Energy contract
DE--AC02--76SF00515}\\
Stanford University, Stanford, CA 94309, USA\\

\vspace{0.7cm}

{\bf  Abstract}
\end{center}

Access to previous results is of paramount importance in the
scientific process. Recent progress in information management focuses
on building e-infrastructures for the optimization of the research
workflow, through both policy-driven and user-pulled dynamics.
For decades, High-Energy Physics (HEP) has pioneered innovative
solutions in the field of information management and dissemination. 
In light of a transforming information environment, it is important
to assess the current usage of information resources by researchers 
and HEP provides a unique test-bed for this assessment.
A survey of about 10\% of practitioners in the field reveals usage
trends and information needs. Community-based services, such as the
pioneering arXiv and SPIRES systems, largely answer the need of the
scientists, with a limited but increasing fraction of younger users
relying on Google. Commercial services offered by publishers or
database vendors are essentially unused in the field.
The survey offers an insight into the most important features that
users require to optimize their research workflow.
These results inform the future evolution of information management in
HEP and, as these researchers are traditionally ``early adopters'' of
innovation in scholarly communication, can inspire developments of
disciplinary repositories serving other communities.
\newpage

\section{Introduction}

High-Energy Physics (HEP), also known as Particle Physics, has a long
record of innovation in scholarly communication. Half a century ago,
theoretical physicists and experimental collaborations mailed to their
peers hundreds, even thousands, of copies of their manuscripts. This occurred
 at the
time of submission to peer-reviewed journals, whose speed in
disseminating scientific information was deemed to be insufficient for
the speed at which the field was evolving~\cite{luisella}. This
practice led to the creation of the first electronic catalog for
gray literature, later evolving into a catalog of the entire subject
literature: the SPIRES database~\cite{luise}.

In the last two decades, crucial innovation in scholarly communication
emerged from the HEP community, ranging from the invention of the
world-wide web at CERN~\cite{weaving} to the inception of arXiv, the first and
archetypal repository~\cite{arxiv}.  The onset of the web gave SPIRES
the honor to be the first web server in America and the first database
 on the web~\cite{kunz}. More recently the HEP community inspired
the development of Invenio, one of the first open-source digital
library software packages~\cite{invenio}, currently used for
repositories in many fields.

Thanks to this suite of user-driven innovations, HEP scholars have
used a variety of dedicated, field-specific ``information
resources''. For many decades these have been run by large research
institutions as a natural evolution of more conventional library
services.  At their inception, these resources often provided unique 
services, or were tailored
specifically to the needs of the HEP community.  Many of these
services still exist and still provide information that cannot be
obtained in any other way.

For many years now almost all journal literature has been
electronically available, the entire web is readily searchable, and
commercial online databases provide metadata about all scientific
literature.  In addition, online services are changing more and more
rapidly as new tools are developed and new ways of interacting with
users evolve. In light of this fast-changing world, it is important to
assess the usage by HEP researchers of HEP-specific information
resources. Such a study serves two purposes: within the field, it
informs on the need for such community-based resources and their real
role in the present internet landscape, inspiring their future
evolution; globally, it provides an in-depth case study of the impact
of discipline-based information resources, as opposed to
institution-based information resources or cross-cutting (commercial)
information platforms. This information is particularly relevant in
light of recent worldwide moves towards self-archiving of research
results at the institutional or disciplinary level, and the need to
effectively incorporate these resources in the research workflow.

A survey of HEP scholars was designed and  
deployed in order to provide a unique insight into their   
information needs and the way their research workflow includes  
information discovery and retrieval.
Its results are presented in this
Article.  The Article first describes the current landscape of HEP
information resources (Section 2), then presents the survey
methodology and the demographics of the respondents (Section 3). Two
sets of results are presented and discussed: the information resources
preferred by HEP researchers (Section 4) and their appreciation of the
relative importance of possible features of information resources
(Section 5). The survey also provides additional information on user
requirements for the future of information resources (Section
6). After the conclusions of the study (Section 7), an Appendix
presents some of the most inspiring free-text answers charting the
future of information provision in this field.

\section{The Landscape of HEP Information Resources}

Several information resources serve the needs of HEP researchers, as
summarized in the following.

\begin{itemize}

  \item {\bf arXiv}~\cite{arXivurl}. arXiv is the archetypal
    repository. It was conceived in 1991 by Paul Ginsparg, then at the Los
    Alamos National Laboratory in New Mexico, and is now hosted at Cornell 
    University in New York. It evolved a four-decade old tradition of HEP
    preprint circulation into an electronic based system, offering all
    scholars a level playing-field from which to access and disseminate
    information. Today arXiv counts nearly 500,000 preprints and 
    has grown outside the field of HEP, becoming
    the reference repository for many diverse disciplines beyond physics,
    from mathematics to some areas of biology. arXiv functions almost
    solely as a repository, with little emphasis on sophisticated
    searching or curation of bibliographic information.  
    
  \item {\bf CDS}~\cite{cdsurl}. The CERN Document Server (CDS) was
    conceived in the late 90s at CERN, the European Organization for
    Nuclear Research in Geneva, Switzerland for the
    management of scientific information at the
    laboratory~\cite{cernlib}. It has a double role. On one side, it is CERN's
    institutional repository, with a mission to archive and disseminate
    CERN results in the fields of experimental and theoretical physics as
    well as accelerator and information technologies. On the other side,
    given the central role of CERN in research in the field, it expanded
    to also offer a gateway to HEP information at large, indexing the
    content of major journals and harvesting full-text from
    many preprint servers, with most of the content coming from
    arXiv. These efforts are more limited in scope and time than
    those at SPIRES, discussed below. CDS is based on the Invenio
    open-source digital library software~\cite{invenio}.  CDS counts about
    1 million records and 500,000 full text-documents, as well as a
    growing multimedia collection.

  \item {\bf SPIRES}~\cite{spiresurl}. SPIRES has provided a
    metadata-only search engine for all literature in the field for over
    30 years. It is hosted at SLAC, the Stanford Linear Accelerator Center
    in California, and jointly compiled together with DESY, the Deutsches
    Elektronen-Synchrotron in Hamburg, Germany, and Fermilab, the Fermi
    National Accelerator Laboratory in Illinois.  SPIRES adds citation
    data, keywords, classifications and authors with their
    institutional affiliations to the basic data that is
    harvested from various sources of physics literature. Today SPIRES has
    grown to include 750,000 records.  SPIRES
    functions primarily as a gateway to all information in the field,
    providing context and consolidation for other data sources. SPIRES
    also provides a corrections and additions capability so that authors
    and users can correct errors they might find. In addition, other
    information of interest to the HEP community is offered at the SPIRES
    site via databases of jobs, conferences, people and institutions~\cite{Kreitz:2003ey}. 
    
\end{itemize}

A fourth, community-based, information resource also serves the HEP community, even
though it was originally designed for the astronomy and astrophysics communities.

\begin{itemize}
  \item {\bf ADS}~\cite{adsurl}. The Astrophysics Data System (ADS) is a
    digital library portal for researchers in astronomy and physics,
    operated by the Smithsonian Astrophysical Observatory at Harvard under
    a NASA grant. It serves as a portal to astronomy and astrophysics
    data as well as bibliographic records. It offers
    highly customizable query forms and gives access to full-text scans of
    much of the astronomical literature which can be browsed or
    searched via a full-text search interface.
 \end{itemize}

The relations that exist between the different HEP information
resources are relevant to understanding the degree of
interoperability and complementarity between them. In addition they help shed
light on both the workflow of HEP researchers and the findings of
the survey. The main relations are as follows:

\begin{itemize}

  \item ADS \& arXiv. ADS maintains a database of all arXiv content
    relevant to HEP, offering additional services such as
    highly customizable e-mail alerts or RSS feeds.

  \item CDS \&  arXiv. CDS carries all arXiv content relevant to HEP,
    with targeted curation effort devoted to matching
    preprints with information on publication reference,
    conference contributions, experimental collaborations, and
    the use of local authority files for author disambiguation.

  \item SPIRES \& arXiv.  Because of their similar histories and
    mostly non-overlapping functions, SPIRES and arXiv could be
    considered as a single system.  arXiv functions as the
    back-end data storage, as well as managing all of the complexities of
    submission.  SPIRES provides a front-end interface, as well as giving
    further context to the arXiv submissions by matching them with
    published literature and adding citation, keywords and other
    data\footnote{This is an oversimplified view, as arXiv.org does
      provide some searching capabilities and, since not all literature is
      submitted to arXiv nor is all arXiv content HEP related, the data sets
      in the two services are not identical.}. Examples of their symbiosis
    include the fact that all of the arXiv content of HEP relevance is indexed in
    SPIRES and arXiv relies on SPIRES for tasks like citation analysis.
    
\end{itemize}

Like virtually everyone else with internet access, HEP scholars also use 
Google~\cite{googleurl} and Google Scholar~\cite{googleScholarurl} 
as information resources. One of the targets of this
study is indeed to assess the penetration of these resources in the HEP
scholarly-communication landscape. It is important to remark that
arXiv and SPIRES have let their content be harvested by Google and then partly
organized in Google Scholar.

\section{Survey Methodology and Demographics}

The data discussed in this Article were obtained by an anonymous
on-line survey widely distributed in the HEP community. The survey ran
for 6 weeks, from April 30, 2007 to June 11, 2007, and collected 2,115
responses. The number of respondents can be compared with the number of
HEP physicists active in 2006, which is about
20,000~\cite{KrauseMele}, or the number of authors who have published
an article listed in SPIRES in the last decade, which is between
30,000 and 40,000, depending on how one handles similar
names. It can be safely concluded that between 5\% and 10\% of the HEP
community participated in the survey. This incredible rate of
participation was further enhanced by the fact that 90\% of the
respondents wrote some optional free-text comments in addition to the
required ``radio-button'' selections, and 73\% responded to optional
lengthier questions. The engagement of the community is further
signified by the fact that about half of the respondents asked to be
informed via e-mail of the results of the poll.

This survey was promoted within the HEP community by e-mailing members
of major experimental collaborations, users of major laboratories and
authors of a major journal of the field: the {\it Journal of High
Energy Physics}. A link to the survey was also distributed for a week
as a heading in the popular daily e-mail alerts of arXiv. Prominent
notices were posted on the CDS and SPIRES websites throughout the
survey and for two weeks on the website of another major journal,
 {\it Physical Review D}. Information on the survey was also
appended for two weeks to the correspondence between the editors and
the authors of a third major journal, the {\it European
Physical Journal C}.

Table 1 presents the distribution of the respondents per country,
which confirms the worldwide character of the HEP community and the
worldwide spread of the survey. The fraction of answers per country
mostly follows the distribution per country of the HEP authorship, as
estimated in References \citen{JHEP} and \citen{cernOpen}, further
confirming the absence of systematic biases in the response to the
survey. The only appreciable trend is a reduced participation to the
survey from Asia. Japan, China, Korea and Taiwan account for about
15\% of HEP authors~\cite{cernOpen}, but they only comprise 5\% of the
respondents to the survey.

Table 2 presents the distribution of the respondents per field of
activity, with theorists accounting for about 60\% of the total
respondents. Table 3 presents the experience of the respondents in the
field, 76\% have been active HEP scholars for 6 years or
more. These data reflect well the demographics of the HEP
community observed in the SPIRES HEPNames database\cite{HEPNames}. The respondents to the survey are heavy users of HEP
information resources: as presented in Table 4, 82\% use
such resources a few times a week or more.

\section{Preferred Systems}

The first question asked in the survey was: {\it Which HEP information
system do you use the most?}  The question, as all other questions
discussed in this section, did not allow multiple choices, offering
``radio buttons'' for arXiv, CDS, Google, Google Scholar and SPIRES,
as well as a free-text box for entering the name of another system.
1\% of the respondents made use of this last possibility, mostly to 
refer to ADS, or to name two systems, typically arXiv and SPIRES,
confirming the perception of these two as a single entity.

The results are presented in Table 5 and Figure 1. Community-based
systems, comprising ADS, arXiv, CDS, SPIRES and local library
services, are the platform of choice for over 91\% of the
respondents. The combination of SPIRES and arXiv represents the vast
majority of this fraction. 9\% of the respondents use Google or Google
Scholar, while commercial systems see a negligible use, around 0.1\%.

An interesting correlation with the seniority of HEP scholars is
observed, whereby Google is the system of choice for 6\% of
scholars active in the field for 10 years or more, but for 22\% of
scholars active in the field for 2 years or less. This trend is
presented in Figure 2. It should be noted that the use of Google or
Google Scholar benefits strongly from the fact that
community-based systems have made their content available for
harvesting. At the same time, Google and Google Scholar also act, as
in many other fields, as a broader alternative to publisher portals,
given that indexing of many publisher websites has taken place in
recent years.

Six further questions were asked to assess the use of different
resources according to the tasks at hand: {\it Which HEP information 
system do you use the most to find...}
\begin{center}
\begin{itemize}
{\it preprints of which you know the reference?}\\
{\it articles of which you know the reference?}\\
{\it preprints on a given subject?}\\  
{\it articles on a given subject?}\\
{\it preprints by a given author?}\\  
{\it articles by a given author?}\\
\end{itemize}
\end{center}
The answers are summarized in Table 5. These more specific questions
reveal that the respondents change their behavior based on the task at
hand, rather than operating only out of loyalty to a particular system.  In
particular their changes in usage seem to match some of the notable
features of the available systems.  As before a trend for a larger usage
of Google by younger scholars is detected for these six questions, 
as summarized in Table 6.

Figure 3 presents an aggregation of the use of community-based
services, Google and commercial systems for the answers concerning
searches for preprints and searches for articles. Again,
community-based services dominate.

As expected, the maximum usage of commercial services, and in
particular publishers' websites, is observed in searches for articles
whose reference is known. However, this number stays remarkably low,
at 4.5\%, confirming that community-based services are the preferred gateway to
information, including the published literature, for HEP scholars.

It is also interesting to remark that arXiv figures as the second
favorite service in searches for articles. This is somewhat
surprising as the site mission is the dissemination of
preprints. However, HEP scholars routinely submit to arXiv an updated
version of their preprints in an author-formatted post-peer-review
version, and therefore make arXiv also a resource for published
literature. Moreover, as the HEP content of arXiv is fully indexed by
SPIRES, and for many users the distinction between the two is
blurred, a user searching for a bibliographic reference in SPIRES,
who clicks on the link to the arXiv version rather than
the publisher version, would be inspired to answer that arXiv is her
system of choice for such a search.

In general, arXiv and SPIRES answer the needs of the vast majority
of users, who do recognize the relative strengths and weaknesses of
these two services as they move back and forth between them according
to the task at hand.  SPIRES is favored for journal literature, while
arXiv increases its direct usage when preprints are
desired.  SPIRES is also more heavily
favored when searching by author, possibly due to its more advanced
author search and better author data.  Subject searching, especially
for published articles, sees a dramatic rise in the use of Google,
possibly because a broader search may be desired, and the
community-based systems do not have the breadth of coverage that
Google has. There are several possible explanations for the lack of
use of commercial systems: few institutions can afford access to them;
if such access exists, most HEP scholars are not aware of such a
possibility; even if they are, these systems do not provide
detailed information specific to HEP users; even if such information 
is available, it is often lost within the ``noise'' of literature from many other fields.

A final question on the preferred systems was:
{\it Which HEP information system do you use the most to find theses?}
The corresponding answers are presented in Table 5 and plotted in Figure 4. 
Unsurprisingly, Google has a larger share than for any other task, at
about 1/3. However the efforts of community services to track and
index theses is still reflected in 2/3 of the users preferring these
services for accessing theses. Commercial systems, again, have a
negligible share, around 0.1\%.

\section{Important Features}

In addition to inquiring about the most heavily used systems for different
tasks, the survey aimed to assess the importance of various
aspects of information resources. Respondents were asked to tag the
importance of 12 features of an information system on a five-step
scale, ranging  from ``not important'' to ``very
important'', these features are:
\begin{center}
\begin{itemize}

\it{Access to full text}\\
\it{Citation analysis}\\
\it{Collaborative tools}\\
\it{Depth of coverage}\\
\it{Keywords and classification}\\
\it{Multimedia content}\\
\it{Personalization}\\
\it{Quality of content}\\
\it{Search accuracy}\\
\it{Speed to find what you want}\\
\it{Submission interface}\\
\it{User friendliness}\\
\end{itemize}
\end{center}
The results are presented in Table 7 and summarized in Figure 5.
Notably, most features are felt to be important. 
9 out of the 12 features were found to be important to over
half of the respondents. Even multimedia content,
the lowest rated feature, was found to be important by 20\% of the
users.

Against this background of important features, access to full-text
stood out clearly as the most valued feature, with only 5 respondents 
of the 1700 who answered this question rating it as not important.
Following close behind full-text access are depth of coverage, quality
of content and search accuracy.  Citation analysis, a feature of many
of the systems listed, was further down the list for most users.  It
was still considered important by most users, but it was clearly a
secondary feature, along with user friendliness.

Even if not all systems offer all these features, the perceived
importance of each feature was found to be mostly independent from the
system most used by the respondents.

The survey included another set of questions, which were clearly
labeled as optional, to further understand which additional features are
considered important. Out of these 21 additional features, 12 are 
particularly relevant and are discussed in the following:
\begin{center}
\begin{itemize}
\bf{Access}\\
\it{Finding theses}\\
\it{Finding conference proceedings}\\
\it{Finding articles cited with a given article}\\
\it{Finding articles citing a given articles}\\
\it{Finding top-cited articles by subject}\\
\bf{Community}\\
\it{Finding conference announcements}\\
\it{Annotating and commenting on documents}\\ 
\it{Directory of authors and affiliations}\\
\it{Retrieving list of publications}\\
\bf{Authorship}\\
\it{Retrieving and exporting article references}\\
\it{Possibility of submitting article revisions}\\
\it{Knowing how often your articles are read}\\
\end{itemize}
\end{center}
The first five features concentrate on the access to information, the
second four are part of a wider service to the community, while the
last three are services tailored to authors.  Respondents were asked
to tag the importance of these features on a five-step scale, ranging
from ``not important'' to ``very important''. The results are
presented in Table 8 and Figure 6. Some of the services which are felt
as moderately or very important by most respondents, such as the possibility of finding all
articles citing a given article and the possibility of submitting a
revised version of an article, are currently offered by SPIRES and
arXiv, respectively. It is interesting to note that two of the other
features which are perceived as moderately or very important by most
respondents, the possibility of finding all articles cited with a given
article and the possibility of knowing how often an article or preprint
is downloaded, are not currently offered by the most widespread services.

\section{Winds of Change}

The survey explicitly inquired about the level of change that HEP
scholars would expect and require from their information
resources: 75\% expected ``some'' to ``a lot of'' change in the next five
years, while only 12\% expected no change\footnote{90\% of respondents answered this question.}.'
To structure this perception of change, respondents were asked to
imagine their ideal information system in five years and
 tag the
importance of 11 possible features on a five-step scale from ``not important'' to ``very important''. These features are:
\begin{center}
\begin{itemize}
\it{Access from your PDA}                               \\
\it{Access to data in figures and tables}               \\
\it{Authoring tools}                                    \\
\it{Centralization}\footnote{The
  full-text of the ``Centralization'' feature was ``Centralization: one single portal to
  all the information''. }  \\

\it{Collaborative tools }                               \\
\it{Connections to fields outside HEP }                  \\
\it{Inclusion of multiple types of documents}           \\
\it{Linked presentation of all instances of a result }\footnote{The full-text of the ``Linked
  presentation'' feature was ``Linked presentation of all instances
  of a result, from notes to theses, from conference slides to articles''.} \\
\it{Multimedia content}                                 \\
\it{Personalization}                                    \\
\it{Recommendation of documents of potential interest } \\
\end{itemize}

\end{center}

The results are presented in Table 9 and Figure 7. While ``modern''
features such
as multimedia content or access from a PDA were not considered
overwhelmingly important, about 90\% of the users tagged three
features as important: the linked presentation of all instances of a
result, the centralization and the access to data in figures and
tables. Immediately following these three is the extension of the
level of service of HEP information systems to 
other, related, disciplines. The last is hardly surprising: SPIRES
has since long bridged the divide towards astrophysics, cosmology and nuclear
physics, following an increased interdisciplinary activity of HEP
scholars.

A final question tried to assess the potential for the implementation
of Web2.0 features to capture user-tagged content. Respondents were
asked: {\it If a simple web interface would show you an article and offer a
  set of categories to which it could belong, how much time would you
  spend in this tagging system to give a service to the community?} 
Of the 90\%  of respondents who answered this question, 19\%
would not spend any time on this system, while 63\% would spend
between five minutes a day and an hour a week. The breakdown of these
answers by the seniority of the respondents is presented in Figure 8.  
There is an immense potential for user-generated, or rather
user-tagged and user-curated, content in the field of information
provision in HEP, as in many other fields of web-based communication.

\section{Conclusions}

The response to the survey was overwhelming, with over 2,000 HEP
scholars, representing about 10\% of the community, answering basic
and long questions, sharing their appreciation and vision for
information management in the field. The large participation is {\it
per se} a result, signifying the engagement of the community with its
information resources.

The main finding of the survey is that community-based services are
overwhelmingly dominant in the research workflow of HEP scholars. 
 Although the popularity of Google increases
with younger researchers, the field-specific utility provided by
these highly-tailored services is perceived as more
relevant.  Commercial systems are virtually unused in the field.

While the various community-based systems have stronger and weaker 
features, users
attach paramount importance to three axes of excellence: access to
full-text, depth of coverage and quality of content.

Future evolution of these systems should be charted by the clear desire of
users for a centralized and coherent presentation of all instances of a
scientific result, with access to data in figures and tables and a
connection to fields outside of HEP. The survey shows that there exists a remarkable
potential for capturing user-tagged content, with a large fraction of
users willing to invest time in such a community service.

The survey collected thousands of free-text answers about the most and
least liked features of current systems and the user requirements for
future evolution of information provision in the field.  While a
detailed study of these additional data is underway, and outside the
scope of this Article, some inspiring answers are distilled in the
Appendix.

The results discussed in this Article confirm the exceptional
situation of the HEP community in the field of scholarly
communication: decades of efforts in developing, maintaining,
populating and curating community-based services enable an efficient
research workflow for HEP scientists and are met by overwhelming user
loyalty.  Scholarly communication is at the dawn of a new era, with the 
onset of institutional repositories and author self-archiving of research results.
In this evolving landscape, could the decades-old
success story of community-based HEP information systems, and their discipline-based 
content aggregation, provide inspiration for scholarly communication in other 
fields?

\section*{Acknowledgments}

First and foremost we wish to thank all HEP scholars who answered this survey,  
 sharing their opinions, suggestions, wishes and constructive criticism  
 on HEP information systems.
We are grateful to our colleagues who shared their insight in the
field of information management, which were crucial in
the preparation of the survey: Catherine Cart, Jocelyne Jerdelet,
Jean-Yves Le Meur, Tibor Simko, Tim Smith, and Jens Vigen at CERN;  Zaven Akopov and Kirsten Sachs at DESY; and Pat Kreitz and Ann Redfield at SLAC. This study would not have reached such a
large audience  without the collaboration of Paul Ginsparg and Simeon Warner
at arXiv, Enrico Balli at SISSA/Medialab, Bob Kelly and Erick Weinberg at APS and Christian
Caron at Springer, who kindly disseminated information about the survey,
and to whom we are indebted.

\newpage

\section*{Appendix: Inspiring Free-Text Answers}

In addition to the results presented above, the survey collected
thousands of free-text answers, inquiring about features of
current systems and their most-desired evolution. A detailed study
of these comments is underway and outside the scope of this Article. However,
it is particularly interesting to distill some of these answers here,
in order to complete the assessment of the engagement of the HEP
community with the systems which serve its information needs and its
expectations for future developments. Some of
the most inspiring free-text answers were along the following lines:

\begin{itemize}
\item Desire for seamless open access to older articles, prior to
  the onset of arXiv in the '90s.
\item Improved full-text search and access to research notes of large
  experimental collaborations. These are a crucial gray-literature channel where large amounts of
  information and details about the results of large experiments transit.
\item Indexing of conference talks and long-term archiving of the corresponding
  slides, beyond the lifetime of conference websites. Interlinking of these
  slides with the corresponding conference proceedings, in preprint
  form with reference to published volumes, and possibly other
  instances describing the results.
\item Use of the HEP information resources as fora for the publication of ancillary material, crucial in the research
  workflow, and in particular:
  \begin{itemize}
    \item numerical data corresponding to tables;
    \item numerical data corresponding to figures;
    \item correlation matrices and additional information beyond these
      presented in tables, to allow an effective re-use of
      scientific results;
    \item fragments of computer code accompanying complex equations in
      articles, to improve the research workflow and reduce the
      possibility of errors;
    \item primary research data in the form of higher-level objects.
  \end{itemize}
\item ``Smarter'' search tools, giving access to articles related to
  articles of interest.
\item Establishment of some new sort of open peer-review, overlaid on arXiv.
\end{itemize}

\newpage
\clearpage

\begin{table}[hbt]
   \begin{center}
     \begin{tabular}{|l|r|l|r|}
       \hline
       Country & Fraction & Country & Fraction \\
       \hline
       United States &27.4\% &Iran      & 0.9\% \\
       Germany       &9.5\%  &Mexico    & 0.9\% \\     
       Italy         &7.7\%  &Australia & 0.8\%\\      
       United Kingdom&6.5\%  &Denmark   & 0.8\%\\      
       CERN          &4.9\%  &Sweden    & 0.8\%\\      
       France        &4.1\%  &Greece    & 0.8\%\\      
       India         &3.4\%  & Portugal & 0.8\%\\      
       Spain         &3.0\%  &Argentina & 0.7\%\\      
       Canada        &2.6\%  &Korea     & 0.7\%\\      
       Brazil        &2.4\%  &Austria   & 0.6\%\\      
       Russia        &2.4\%  &Poland    & 0.6\%\\      
       Switzerland   &2.2\%  &Chile     & 0.5\%\\      
       China         &2.1\%  &Finland   & 0.5\%\\      
       Japan         &1.7\%  &Taiwan    & 0.5\%\\      
       Israel        &1.5\%  &Czech Republic&0.3\%\\   
       Netherlands   &1.2\%  &Norway    & 0.2\%\\      
       Belgium       &1.1\%  &Hungary   & 0.2\%\\      
       Turkey        &0.9\%  &Others    & 4.8\%\\      
       \hline
     \end{tabular}
     \caption{Distribution of answers per country. Users based at CERN
     were asked to indicate ``CERN'' and not ``Switzerland''. 97\%
     of respondents answered this question.}
   \end{center}
   \label{tab:1}
\end{table}

\begin{table}[hbt]
   \begin{center}
     \begin{tabular}{|l|r|}
       \hline
Field of activity &  Fraction\\
\hline
Theory          &61.3\%\\
Experiment      &22.2\%\\
Software        &5.5\%\\
Instrumentation &3.5\%\\
Accelerators    &2.7\%\\
Engineering     &1.3\%\\
Others          &3.5\%\\
\hline
     \end{tabular}
     \caption{Field of activity of respondents to the
       survey. 96\% of respondents answered this question.}
   \end{center}
   \label{tab:2}
\end{table}

\begin{table}[hbt]
   \begin{center}
     \begin{tabular}{|l|r|}
       \hline
       How long have you & \\
       used HEP search engines?& Fraction\\
       \hline
       $>$10 years&45.9\%\\
       6-10 years&30.0\%\\
       3-5 years&18.7\%\\
       0-2 years&5.4\%\\
       \hline
     \end{tabular}
     \caption{Experience of respondents. 95\% of respondents
       answered this question.}
   \end{center}
   \label{tab:3}
\end{table}

\begin{table}[hbt]
   \begin{center}
     \begin{tabular}{|l|r|}
       \hline
How frequently do you &\\
use HEP search engines?&Fraction\\
\hline
Every day&57.0\%\\
A few times per week&25.6\%\\
Once a week&5.5\%\\
A few times per month&7.4\%\\
Once a month&2.0\%\\
A few times a year&2.5\%\\
\hline
     \end{tabular}
     \caption{Frequency of use of HEP search engines. 95\% of respondents
     answered this question.}
   \end{center}
   \label{tab:4}
\end{table}

\clearpage

\begin{sidewaystable}
  \begin{center}
    \begin{tabular}{|l|r|r|r|r|r|r|r|r|r||r|}
\cline{2-10}
\multicolumn{1}{l|}{} & \multicolumn{5}{c|}{Community Services} & \multicolumn{2}{c|}{Google}&  \multicolumn{2}{c||}{Commercial services} &    \multicolumn{1}{l}{}\\  
\cline{2-11}
\multicolumn{1}{l|}{}    & &&&&Library&&Google&Publisher&Commercial& Fraction\\
\multicolumn{1}{l|}{}         & SPIRES&arXiv&CDS&ADS&services&Google&Scholar&website&databases&of answers\\
\hline
    \multicolumn{11}{|l|}{Which HEP information system do you use the most?}\\
\hline
&48.2\%&39.7\%&2.6\%&0.7\%&0.2\%&7.8\%&0.7\%&0.0\%&0.1\%&99\%\\
\hline
    \multicolumn{11}{|l|}{Which HEP information system do you use the most to find...}\\
\hline
preprints (known reference)?&46.0\%&45.2\%&2.9\%&0.2\%&0.2\%&4.2\%&0.9\%&0.3\%&0.1\%&97\%\\
articles (known reference)?&60.2\%&20.6\%&3.1\%&2.1\%&1.4\%&6.1\%&1.5\%&4.5\%&0.5\%&96\%\\
preprints on a given subject?&49.6\%&34.2\%&3.2\%&0.8\%&0.0\%&9.7\%&2.4\%&0.1\%&0.0\%&96\%\\
articles on a given subject?&55.2\%&19.4\%&2.3\%&1.7\%&0.3\%&16.0\%&3.9\%&0.5\%&0.7\%&95\%\\
preprints by a given author?&67.1\%&24.0\%&2.8\%&0.5\%&0.1\%&4.2\%&1.0\%&0.0\%&0.2\%&96\%\\
articles by a given author?&74.0\%&12.1\%&2.4\%&1.8\%&0.2\%&6.7\%&1.7\%&0.3\%&0.8\%&96\%\\
theses?&48.8\%&10.9\%&4.3\%&0.9\%&2.4\%&27.2\%&5.4\%&0.0\%&0.1\%&83\%\\
\hline
    \end{tabular}
    \caption{Favorite information systems in general, first row, and
      for specific needs. The last column summarizes the fraction of
      respondents who answered these questions.}
  \end{center}
   \label{tab:5}
\end{sidewaystable}

\clearpage

\begin{table}[hbt]
   \begin{center}
     \begin{tabular}{|l|r|r|}
       \cline{2-3}
 \multicolumn{1}{c|}{}  &  \multicolumn{2}{c|}{Fraction of users of}\\
\cline{1-1}
Which system do you use      &  \multicolumn{2}{c|}{Google and Google Scholar}\\
\cline{2-3}
 the most...        & $>10$ years & $<2$ years \\
\hline
in absolute                      &6.0\%&22.1\%\\  
for preprints (known reference)? &3.2\%&15.4\%\\ 
for articles (known reference)?  &5.4\%&16.2\%\\ 
for preprints on a given subject?&9.3\%&29.5\%\\ 
for articles on a given subject? &17.5\%&34.4\%\\
for preprints by a given author? &3.1\%&17.0\%\\ 
for articles by a given author?  &5.9\%&20.1\%\\ 
for theses?                      &30.0\%&41.0\%\\
\hline
     \end{tabular}
     \caption{Penetration of Google and Google Scholar as information
       resources in HEP as a function of the seniority of the scholars.}
   \end{center}
   \label{tab:6}
\end{table}

\clearpage

\begin{sidewaystable}
  \begin{center}
    \begin{tabular}{|l|r|r|r|r|r||r|}
\cline{2-7}
\multicolumn{1}{l|}{}&Very& Moderately& Somewhat & Slightly & Not & Fraction \\
\cline{1-1}
Feature&important&important&important&important&important& of answers\\
\hline
Access to full text&85.8\%&11.9\%&1.9\%&0.1\%&0.2\%&83\%\\
Depth of coverage&72.7\%&21.8\%&4.9\%&0.5\%&0.1\%&82\%\\
Quality of content&63.3\%&27.7\%&7.7\%&0.7\%&0.6\%&82\%\\
Search accuracy&59.5\%&33.9\%&6.1\%&0.3\%&0.2\%&83\%\\
Speed to find what you want&51.2\%&35.4\%&11.4\%&1.5\%&0.5\%&83\%\\
User friendliness&48.6\%&36.8\%&12.3\%&1.5\%&0.9\%&83\%\\
Citation analysis&35.3\%&31.1\%&21.8\%&8.3\%&3.5\%&82\%\\
Submission interface&28.3\%&37.3\%&27.7\%&5.0\%&1.7\%&81\%\\
Keywords and classification&21.4\%&28.0\%&29.1\%&13.5\%&8.0\%&80\%\\
Collaborative tools&10.9\%&17.1\%&34.2\%&19.8\%&18.0\%&75\%\\
Personalization&6.5\%&13.9\%&28.7\%&26.7\%&24.3\%&79\%\\
Multimedia content&6.0\%&14.4\%&29.7\%&25.0\%&24.9\%&79\%\\
\hline
    \end{tabular}
    \caption{Perceived importance of single features of a HEP
      information system. The last column summarizes the fraction of
      respondents who answered these questions.}
  \end{center}
   \label{tab:7}
\end{sidewaystable}

\clearpage

\begin{sidewaystable}
  \begin{center}
    \begin{tabular}{|l|r|r|r|r|r||r|}
\cline{2-7}
\multicolumn{1}{l|}{}&Very& Moderately& Somewhat & Slightly & Not & Fraction \\
\cline{1-1}
Feature&important&important&important&important&important& of answers\\
\hline
Finding articles citing a given article    &49.9\%&30.0\%&13.1\%&4.7\%&2.3\%  &69.5\%  \\
Finding articles cited with a given article&44.1\%&34.4\%&15.7\%&4.3\%&1.6\%  &69.8\%  \\
Finding conference proceedings             &32.4\%&36.9\%&21.0\%&6.7\%&3.0\%  &69.6\%  \\
Finding top-cited articles by subject      &25.0\%&31.0\%&25.2\%&13.5\%&5.3\% &68.3\%  \\
Finding theses                             &23.3\%&31.3\%&22.3\%&15.0\%&8.2\% &69.1\%  \\
\hline
Retrieving list of publications            &24.8\%&30.2\%&26.2\%&13.4\%&5.4\% &66.3\%  \\
Finding conference announcements           &23.3\%&35.2\%&23.3\%&12.5\%&5.6\% &69.1\%  \\
Directory of authors and affiliations      &12.0\%&23.0\%&32.5\%&19.7\%&12.8\%&65.7\%  \\
Annotating and commenting on documents     &8.0\%&16.3\%&26.3\%&25.7\%&23.9\% &65.0\%  \\
\hline
Possibility of submitting article revisions&45.7\%&33.0\%&15.4\%&3.8\%&2.1\%  &67.0\%  \\
Knowing how often your articles are read   &31.7\%&35.3\%&22.2\%&7.6\%&3.2\%  &67.5\%  \\
Retrieving and exporting article references&22.3\%&31.2\%&27.0\%&13.3\%&6.2\% &66.3\%  \\
\hline
    \end{tabular}
    \caption{Perceived importance of additional features of a HEP
information system. The first five features concentrate on the access to
information, the second four are part of a wider service to the
community while the last three are services tailored to authors. The
last column summarizes the fraction of respondents who answered these
questions.}
  \end{center}
   \label{tab:8}
\end{sidewaystable}

\clearpage

\begin{sidewaystable}
  \begin{center}
    \begin{tabular}{|l|r|r|r|r|r||r|}
\cline{2-7}
\multicolumn{1}{l|}{}&Very& Moderately& Somewhat & Slightly & Not & Fraction \\
\cline{1-1}
Feature&important&important&important&important&important& of answers\\
\hline
Access to data in figures and tables                &34.6\%&35.5\%&17.7\%&8.2\%&4.0\%   &88.1\%  \\
Centralization                                      &34.2\%&33.2\%&20.7\%&7.8\%&4.1\%   &87.5\%  \\
Linked presentation of all instances of a result    &30.3\%&38.7\%&21.2\%&6.6\%&3.1\%   &87.7\%  \\
Connections to fields outside HEP                   &26.2\%&28.1\%&25.3\%&13.7\%&6.7\%  &88.1\%  \\
Inclusion of multiple types of documents            &19.5\%&29.8\%&27.2\%&13.2\%&10.3\% &87.1\%  \\
Recommendation of documents of potential interest   &15.4\%&30.4\%&28.0\%&16.2\%&10.0\% &87.6\%  \\
Authoring tools                                     &12.8\%&25.6\%&32.5\%&16.9\%&12.2\% &84.5\%  \\
Collaborative tools                                 &11.5\%&23.6\%&33.2\%&19.0\%&12.7\% &83.7\%  \\
Access from your PDA                                &8.9\%&15.8\%&22.9\%&23.3\%&29.0\%  &86.6\%  \\
Multimedia content                                  &7.4\%&15.3\%&28.8\%&25.9\%&22.7\%  &86.0\%  \\
Personalization                                     &6.7\%&18.4\%&32.7\%&26.3\%&15.9\%  &85.4\%  \\
\hline
    \end{tabular}
    \caption{Perceived importance of future features of a HEP
      information system. The last column summarizes the fraction of
      respondents who answered these questions.}
  \end{center}
   \label{tab:9}
\end{sidewaystable}

\newpage

\begin{figure}[p]
  \begin{center}
    \includegraphics[width=\textwidth]{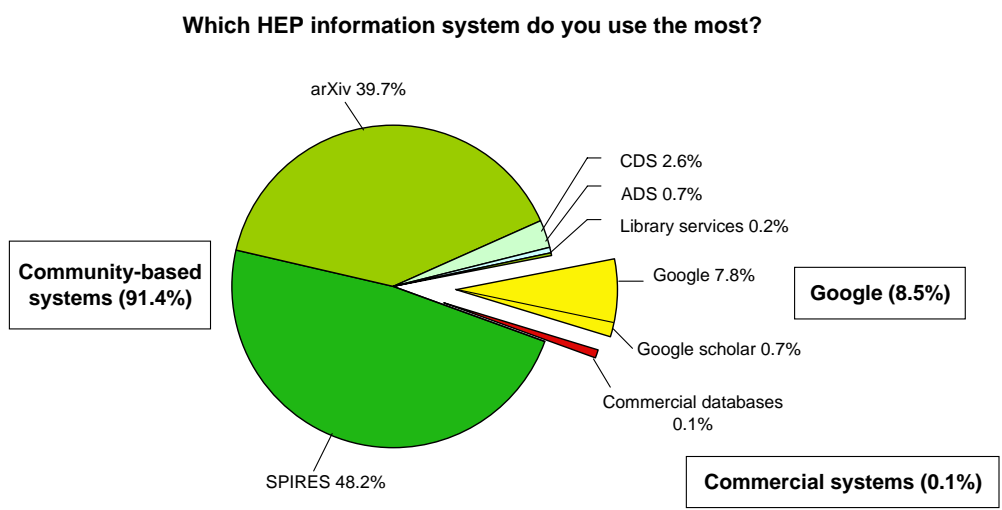}
    \caption{Favorite information resources for HEP scholars. The slice
      corresponding to commercial systems is enlarged for increased
    visibility.}
  \label{fig:1}
  \end{center}
\end{figure}

\clearpage

\begin{figure}[h]
  \begin{center}
   \includegraphics[width=\textwidth]{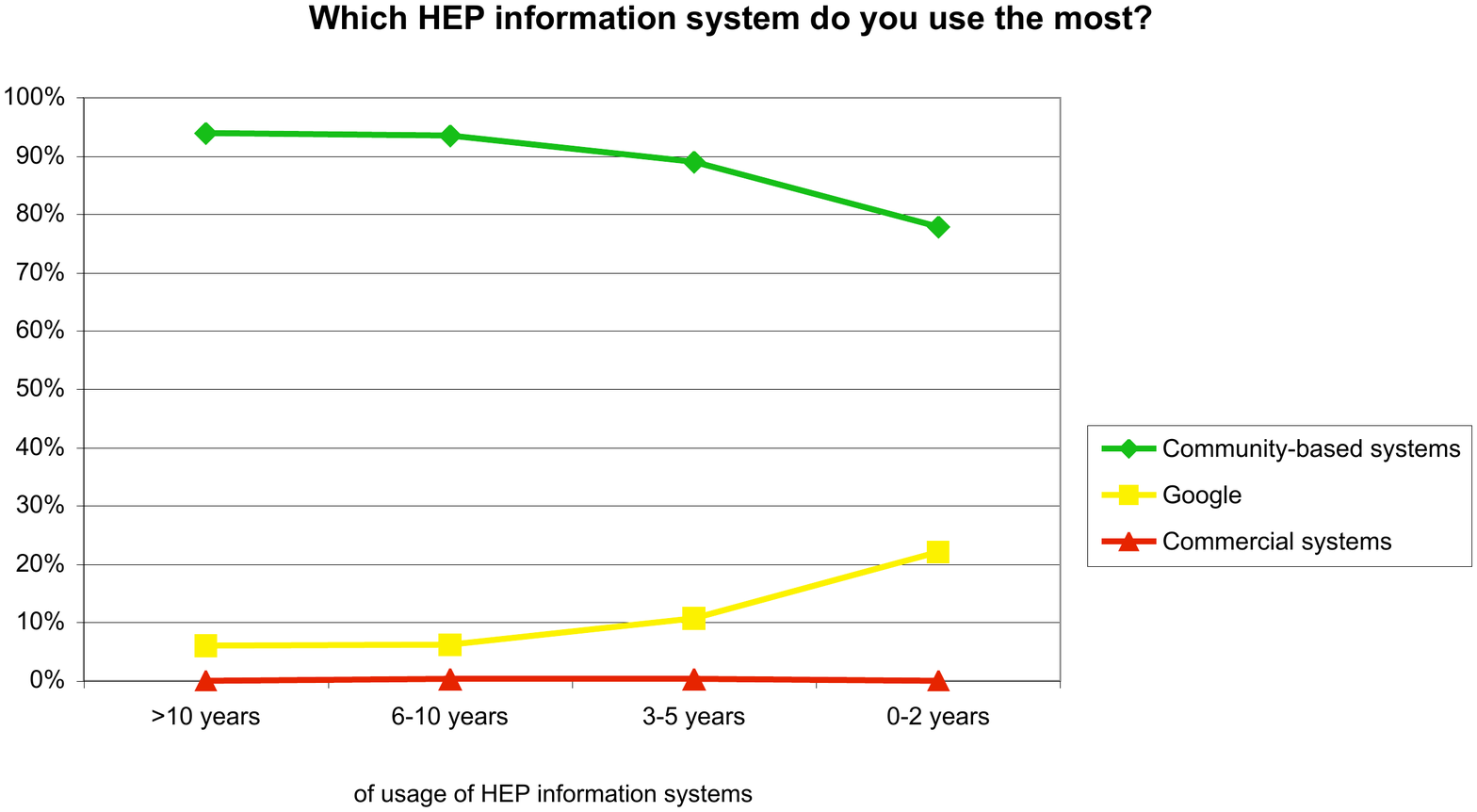}
    \caption{Categories of information resources used by HEP scholars
      as a function of their seniority in the field.}
  \label{fig:2}
  \end{center}
\end{figure}

\clearpage

\begin{figure}[h]
  \begin{center}
    \begin{tabular}{c}
    \mbox{\includegraphics[width=0.75\textwidth]{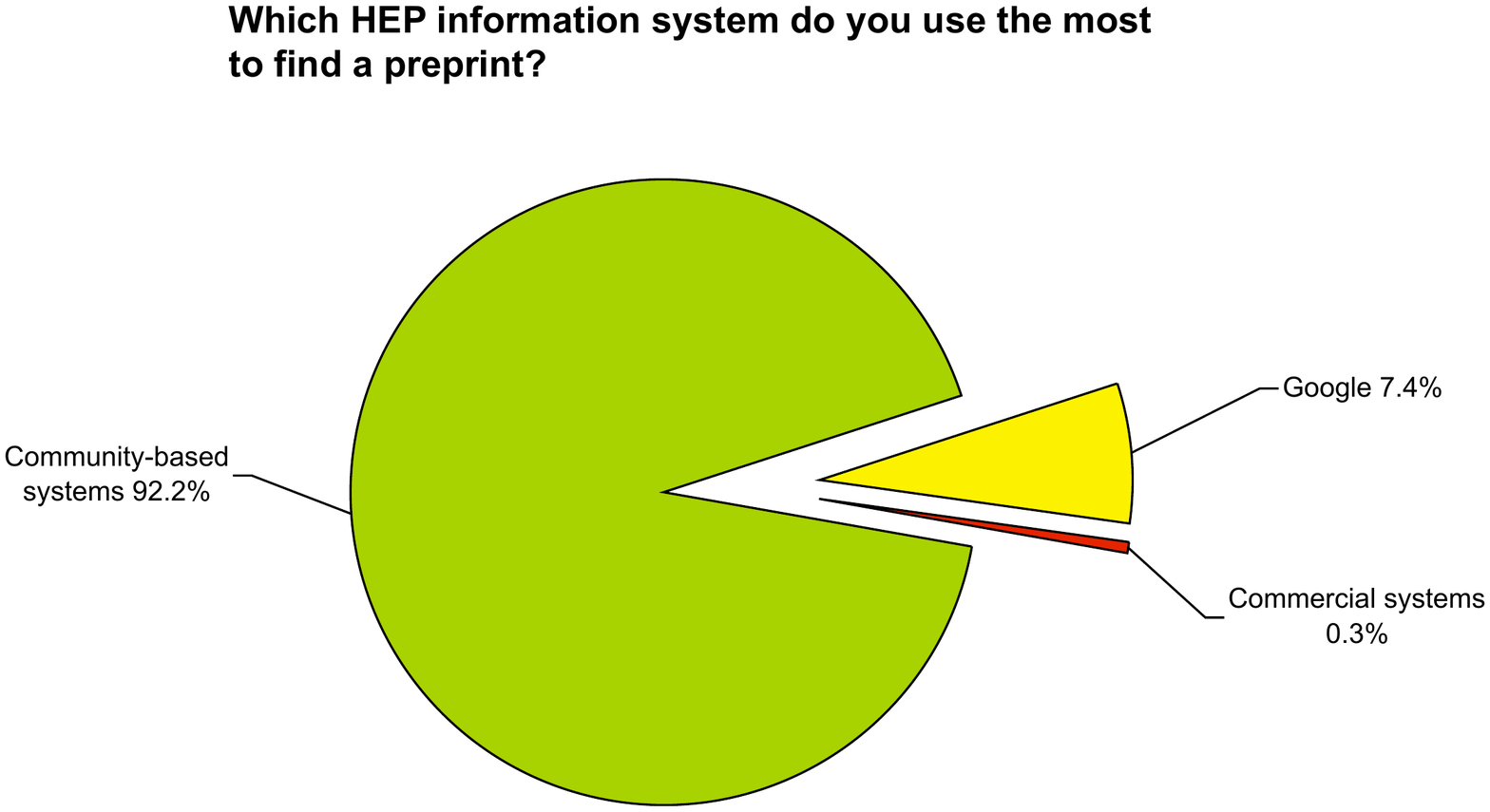}}\\
    \mbox{\includegraphics[width=0.75\textwidth]{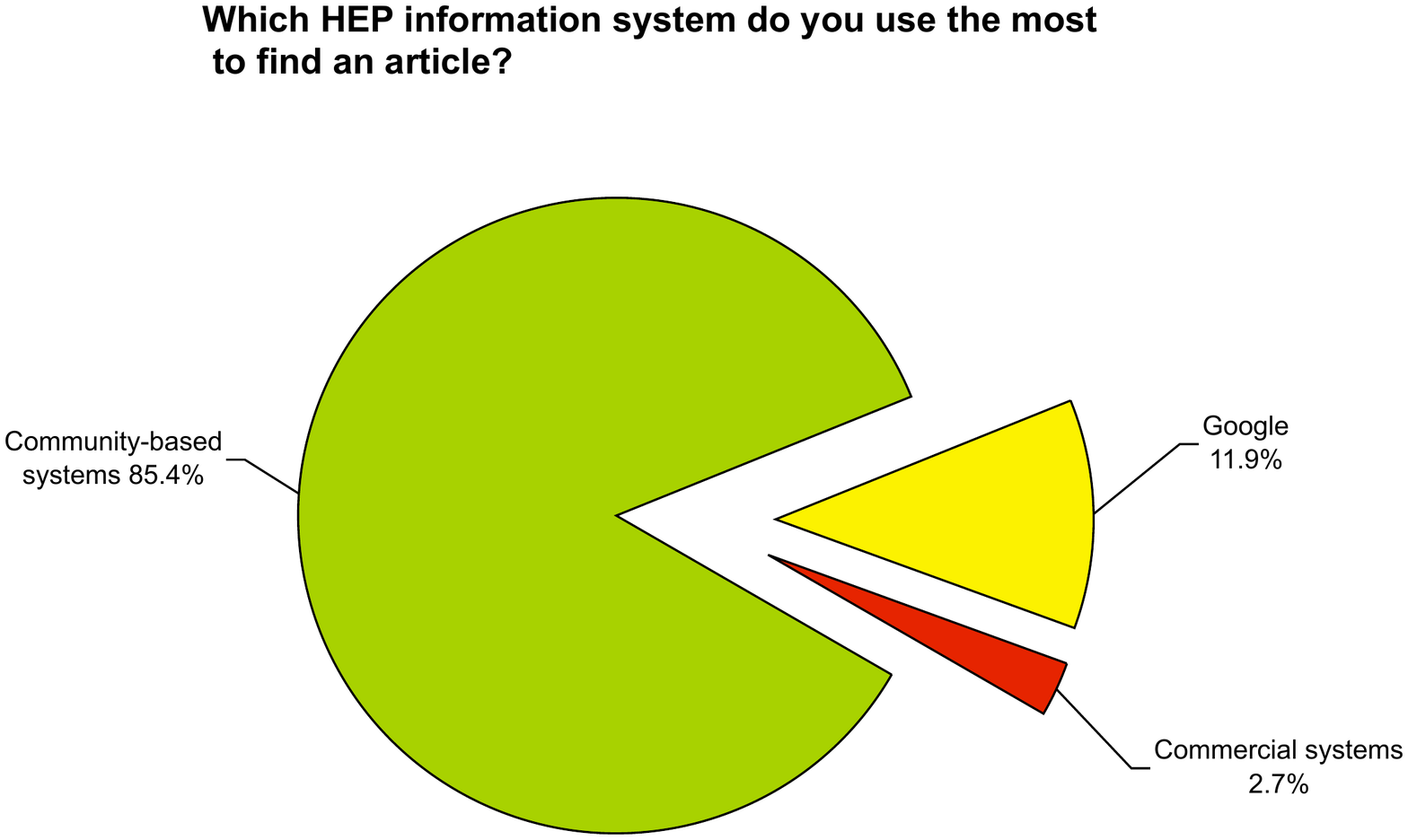}}
    \end{tabular}
    \caption{Type of information resource most used by HEP scholars to
    access information in the form of preprints or published articles.}
  \label{fig:3}
  \end{center}
\end{figure}

\clearpage

\begin{figure}[h]
  \begin{center}
   \includegraphics[width=\textwidth]{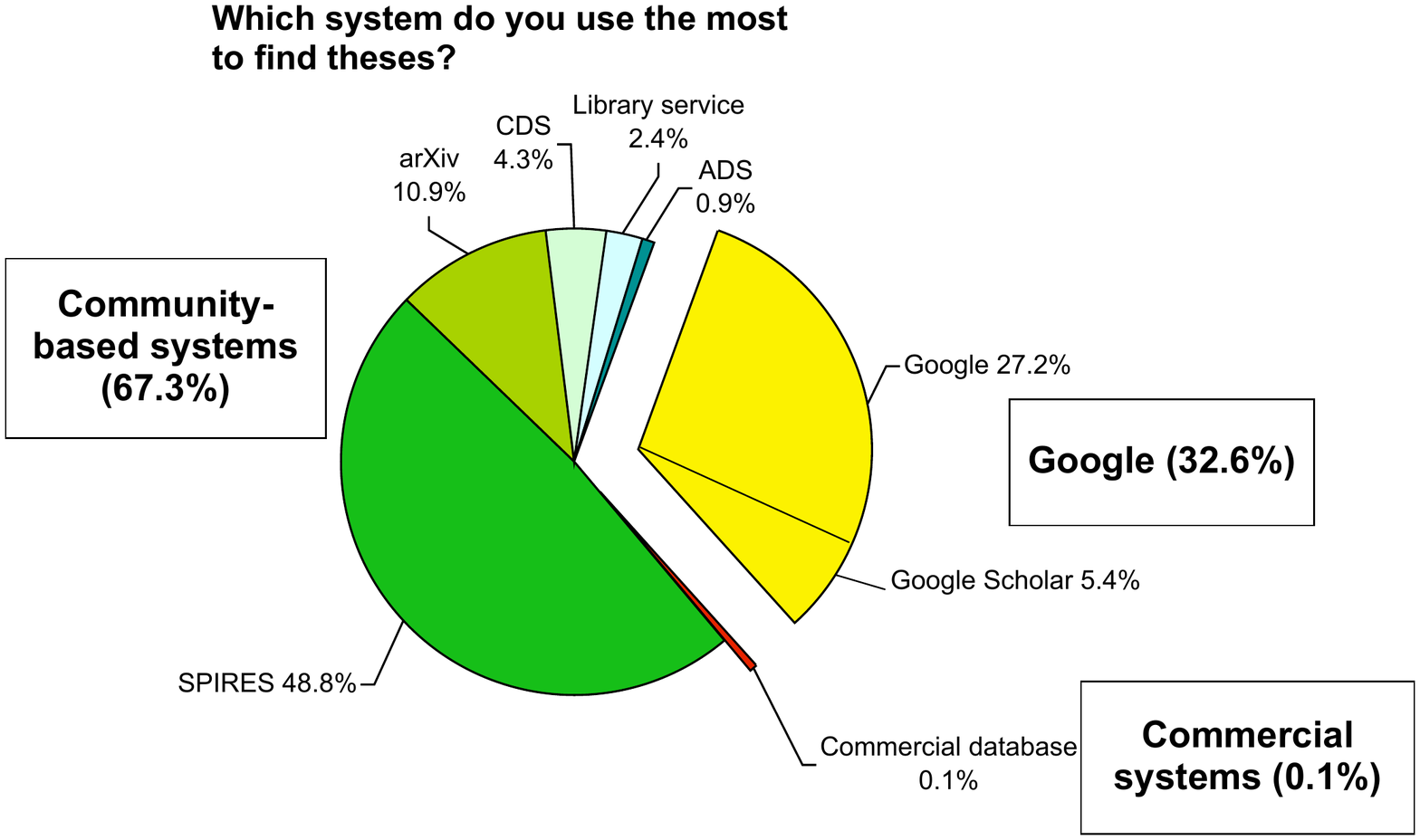}
    \caption{Information resources most used by HEP scholars to search
      for theses.  The slice corresponding to commercial systems is enlarged
      for increased visibility.}
  \label{fig:4}
  \end{center}
\end{figure}

\clearpage

\begin{sidewaysfigure}[h]
  \begin{center}
    \includegraphics[height=0.75\textwidth]{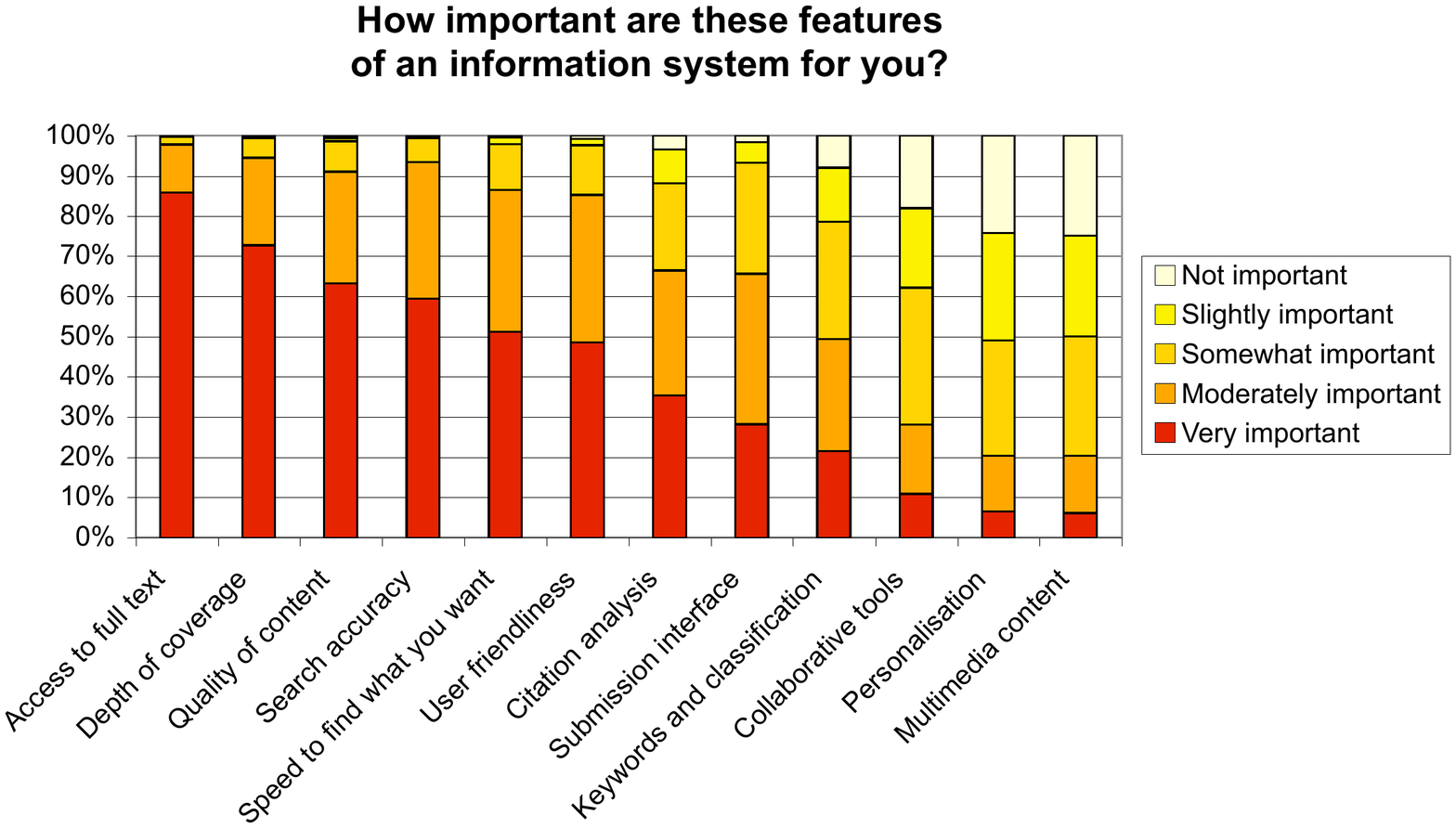}
    \caption{Perceived importance of features of HEP information resources.}
  \label{fig:5}
  \end{center}
\end{sidewaysfigure}

\clearpage

\begin{sidewaysfigure}[h]
  \begin{center}
    \includegraphics[height=0.75\textwidth]{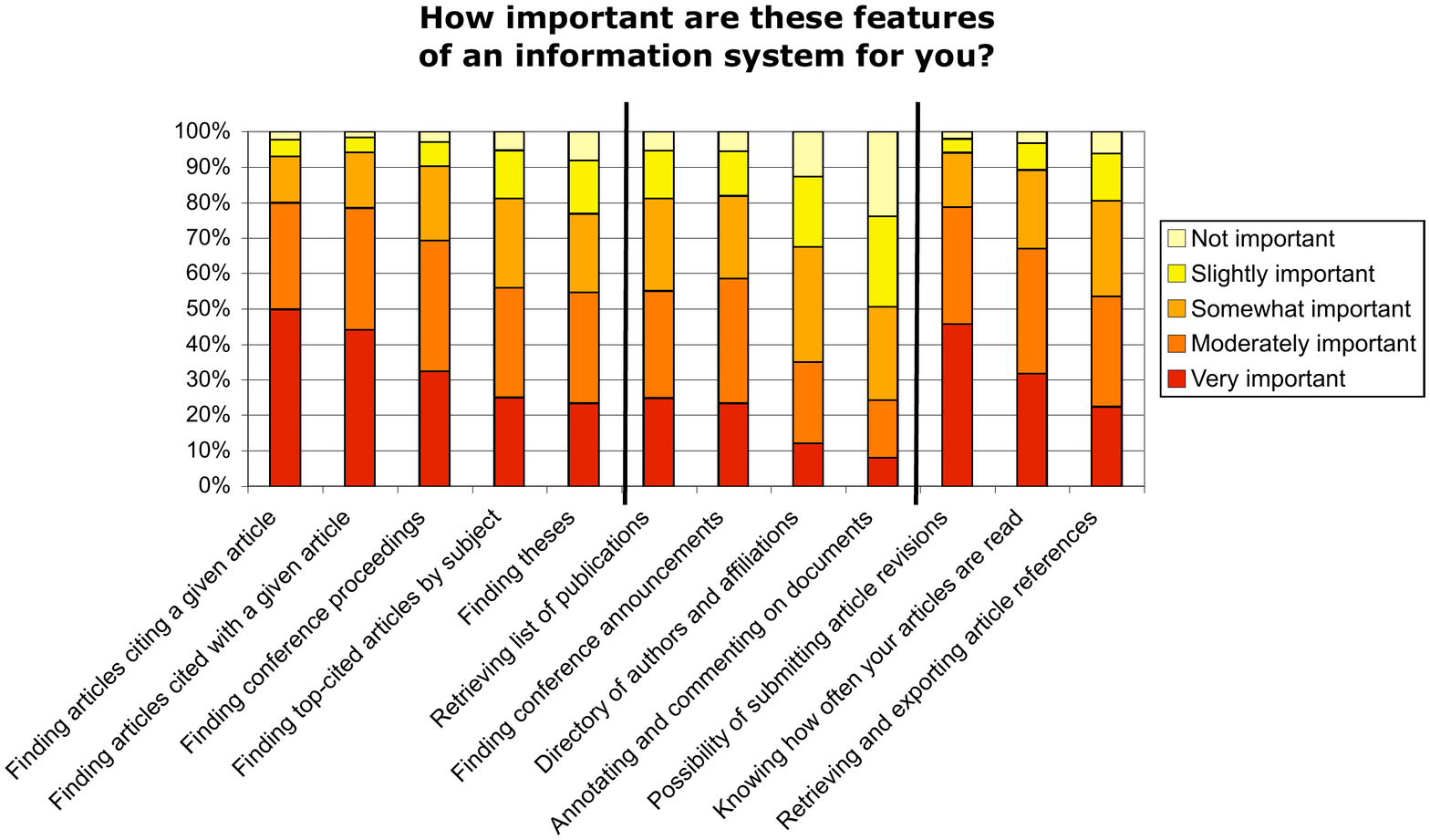}
    \caption{Perceived importance of additional features of HEP information
      resources. The first five features concentrate on the access to
      information, the second four are part of a wider service to the
      community while the last three are services tailored to authors.}
    \label{fig:6}
  \end{center}
\end{sidewaysfigure}

\clearpage

\begin{sidewaysfigure}[h]
  \begin{center}
    \includegraphics[height=0.75\textwidth]{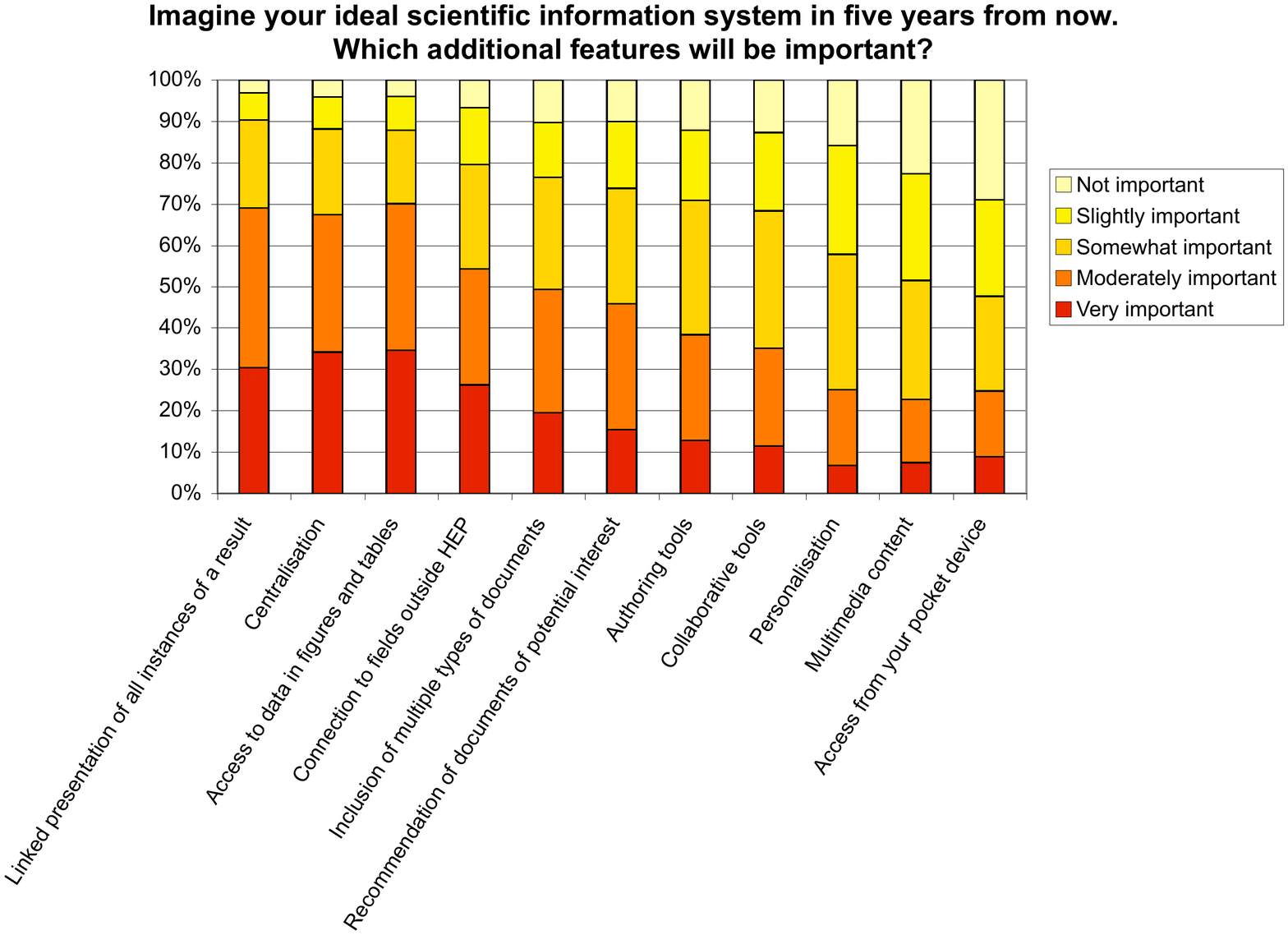}
    \caption{Perceived importance of future features of HEP
      information resources. }
  \label{fig:7}
  \end{center}
\end{sidewaysfigure}

\clearpage 

\begin{figure}[h]
  \begin{center}
    \includegraphics[width=\textwidth]{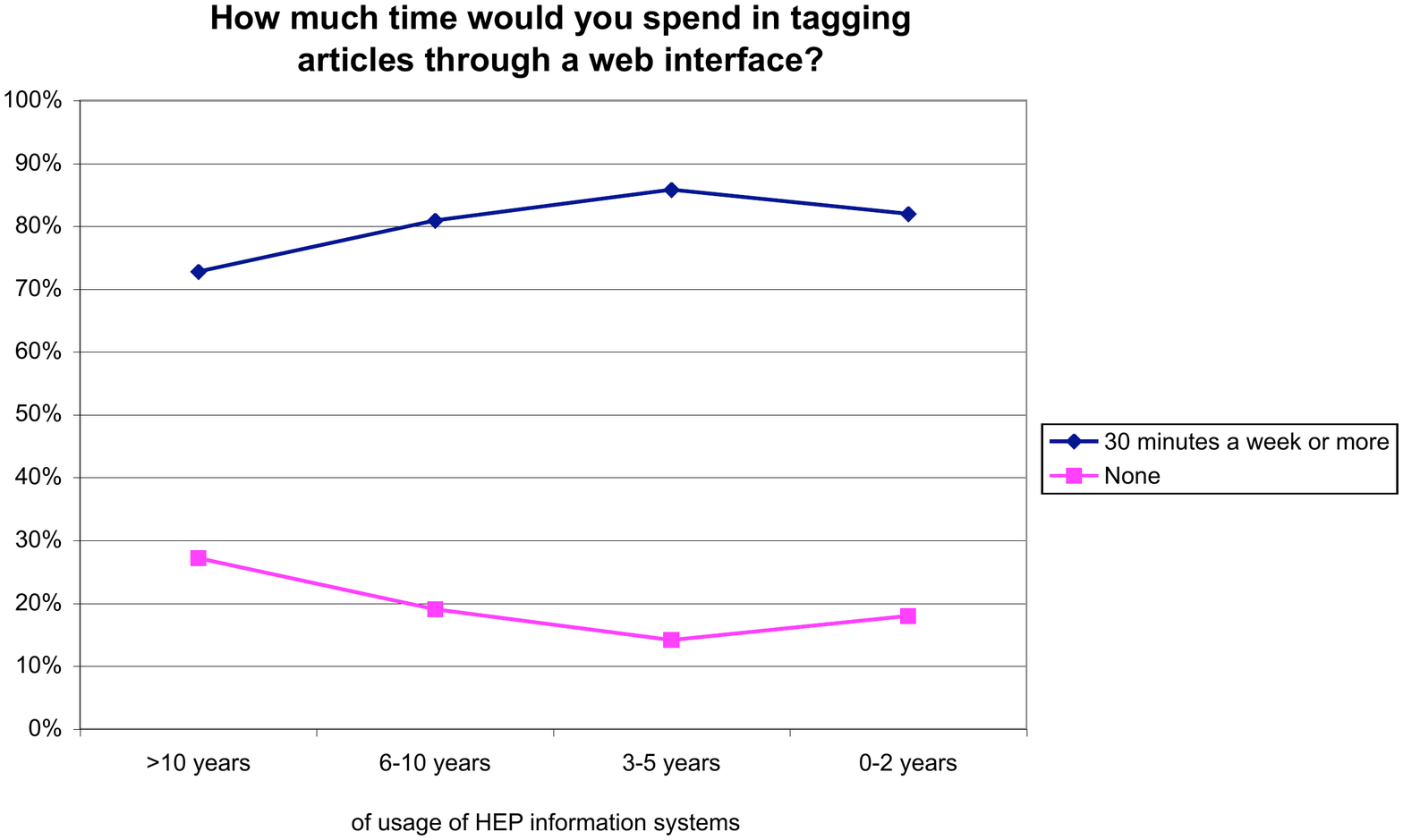}
    \caption{Interest in participating in user-tagging of content as a
      function of the seniority of respondents.}
  \label{fig:8}
  \end{center}
\end{figure}

\clearpage

\end{document}